\documentclass[a4paper,conference]{IEEEtran}
\IEEEoverridecommandlockouts
\usepackage{stfloats,color}
\usepackage{amsfonts}
\usepackage{setspace}
\usepackage{adjustbox}
\usepackage{ragged2e}
\usepackage{cite}
\usepackage{bbm}

\usepackage{filecontents}
\usepackage{array}

\usepackage{mdwmath}
\usepackage{multirow}
\usepackage{makecell}
\usepackage{times}
\usepackage{diagbox}
\usepackage{epsfig}
\usepackage{latexsym}
\usepackage{epstopdf}
\usepackage{verbatim}
\usepackage{units}
\usepackage{amsthm}
\usepackage{mdwlist}
\usepackage{esvect}
\usepackage{tabularx,colortbl}

\usepackage[unicode=true, linktocpage, linkbordercolor={0.5 0.5 1}, citebordercolor={0.5 1 0.5}, linkcolor=blue]{hyperref}

\begin{document}

\title{Epileptic Seizure Prediction: A Semi-Dilated Convolutional Neural Network Architecture}






\author{\IEEEauthorblockN{
Ramy Hussein\IEEEauthorrefmark{1},
Soojin Lee \IEEEauthorrefmark{2}\IEEEauthorrefmark{3},
Rabab Ward\IEEEauthorrefmark{4} and
Martin J. McKeown\IEEEauthorrefmark{2} 
}
\IEEEauthorblockA{\IEEEauthorrefmark{1}Center for Artificial Intelligence in Medicine \& Imaging, Stanford University}
\IEEEauthorblockA{\IEEEauthorrefmark{2}Pacific Parkinson's Research Centre, University of British Columbia}
\IEEEauthorblockA{\IEEEauthorrefmark{3}Wellcome Centre for Integrative Neuroimaging, University of Oxford}	
\IEEEauthorblockA{\IEEEauthorrefmark{4}Electrical and Computer Engineering, University of British Columbia}
}

\maketitle

\begin{abstract}

Accurate prediction of epileptic seizures has remained elusive, despite the many advances in machine learning and time-series classification.  In this work, we develop a convolutional network module that exploits Electroencephalogram (EEG) scalograms to distinguish between the \textit{pre-seizure} and \textit{normal} brain activities. Since these scalograms have rectangular image shapes with many more temporal bins than spectral bins, the presented module uses ``semi-dilated convolutions'' to create a proportional non-square receptive field. The proposed semi-dilated convolutions support exponential expansion of the receptive field over the long dimension (image width, i.e. time) while maintaining high resolution over the short dimension (image height, i.e., frequency). The proposed architecture comprises a set of co-operative semi-dilated convolutional blocks, each block has a stack of parallel semi-dilated convolutional modules with different dilation rates. Results show that our proposed solution outperforms the state-of-the-art methods, achieving seizure prediction sensitivity scores of 88.45\% and 89.52\% for the American Epilepsy Society and Melbourne University EEG datasets, respectively.

\end{abstract}



\IEEEpeerreviewmaketitle

\section{Introduction}


Epilepsy is a chronic neurological disorder characterized by recurrent and unprovoked seizures \cite{Stafstrom2015}. It is the second most common neurological condition that affects 1\% of the population, with approximately 75\% of cases starting in childhood \cite{Sperling2004}. The mainstay of epilepsy treatment is pharmacotherapy, with anti-epileptic medications effective in controlling seizures in around 70\% of people \cite{French2007}. Medications may control, but do not cure epilepsy, and can result in undesirable side-effects, some of which are fatal \cite{Ortinski2004}. Approximately one-third of epileptic patients are drug-resistant and continue to have seizures despite adequate anti-epileptic drugs \cite{Dalic2016}. In such cases, the timely prediction of the impending seizure would be extremely beneficial as it (i) helps epileptic patients take precautionary measures against possible seizure-related injuries and  (ii) enables employment of seizure intervention solutions that could potentially control or abort the upcoming seizure \cite{Cook2013prediction}.


Seizure prediction systems typically use intracranial or scalp Electroencephalogram (EEG) data, as it has the temporal resolution to detect electrical abnormalities in brain activities \cite{Smith2005}. A seizure prediction algorithm starts with extracting the discriminative EEG features from the time domain, frequency domain, time-frequency domain, or a combination of multiple domains \cite{hussein2019scalp}. The extracted features are then fed into a decision-making classifier that can distinguish between preictal (before seizure) and interictal (\textit{i.e.}, normal) brain states. Many linear and non-linear features of EEG have been used for predicting epileptic seizures. The most commonly used linear features include the power of EEG rhythms (delta, theta, alpha, beta, gamma) and statistical measures of EEG signals (\textit{e.g.}, mean, variance, skewness, kurtosis, etc.) \cite{mormann2005predictability, Rasekhi2013, Brinkmann2016, Zhang2016, Kuhlmann2018}. Non-linear features include entropy, correlation, dynamical similarity, and also Lyapunov exponent \cite{mormann2005predictability, Brinkmann2016, Kuhlmann2018}. However, such domain-based and hand-crafted features can not maintain reliable prediction performance in clinical settings \cite{Usman2017}.

Because of the limitations of EEG feature engineering methods, a generalized seizure prediction algorithm that can perform automatic feature extraction would be extremely beneficial \cite{hussein2019human}. Convolutional neural network (CNN) is one of the promising deep learning architectures that can handle raw data and automatically extract the distinguishable EEG features needed for accurate seizure prediction. 
CNN was first used for seizure prediction by Truong \textit{et al.} \cite{Truong2018}, where a two-dimensional CNN model was applied to the short-time Fourier transform of EEG signals to learn the distinctive features that can distinguish preictal data from interictal ones. In \cite{Zhang2019}, CNN was also used together with a feature extractor employing wavelet packet decomposition and common spatial pattern to elicit the discriminatory features from both time and frequency domains. Further, several CNN architectures were also presented in \cite{Eberlein2018, Korshunova2018, Liu2019} to mitigate multiple views of EEG by integrating the time domain and frequency domain as separable inputs. While the seizure prediction results are strikingly good, there is still plenty of room for improvement by employing adequate CNN architectures.

In this work, we propose a generalized seizure prediction algorithm based on a novel convolutional module named ``semi-dilated convolution''. The main contributions of this work are summarized as follows: (1) we propose to use an efficient pre-processing strategy that transforms the time-series EEG data into image-like representations named ``scalogram'', (2) inspired by dilated convolution, we develop a novel convolutional module named ``semi-dilated convolution'' that can better leverage the geometry of nonsquare-shape images such as scalograms, and (3) we present a co-operative neural network architecture named ``semi-dilated convolutional network (SDCN)'' for automated and efficient EEG feature learning and classification. Our proposed seizure prediction algorithm outperforms the baseline methods on two popular large-scale EEG datasets recorded from dogs and humans.    


\section{Materials and Related Work}

\subsection{Seizure prediction using canine and human invasive EEG}
\label{sec2-1}

\textbf{Dataset:} American Epilepsy Society Seizure Prediction Challenge \cite{Brinkmann2016} --- invasive EEG (iEEG) was recorded at 400 Hz using an ambulatory system with 16 electrodes from eight dogs with naturally occurring epilepsy. Of the eight dogs, five produced data with high quality and an adequate number of seizures. The dataset also includes iEEG recorded from two epileptic patients with drug-resistant epilepsy. Both patients were females and they were 70 and 48 years old in the time of iEEG being recorded. The iEEG data of the first patient was recorded for 71.3 hours and that of the second patient was recorded for 158.5 hours, using 3 $\times$ 8-contact subdural electrodes. All iEEG data were provided in 10-minute data clips: preictal clips were selected from the 66 minutes prior to seizures, and interictal clips were extracted similarly in six 10-minute clips beginning from a randomly selected time point that was at least 1 week before any seizure.

\textbf{Related Work} --- The American Epilepsy Society launched a seizure prediction competition on Kaggle releasing the iEEG data to classify the interictal and preictal clips. More than 500 teams participated in the competition and the winning team achieved an average area under the curve (AUC) score of 0.90 and 0.84 on the public and private test sets, respectively \cite{Brinkmann2016}. The algorithms of other top participating teams are described and summarized in Table~\ref{Tab_AES}. They are ranked based on their seizure prediction performance on the private leaderboard, \textit{i.e.}, private test set. Most of the top finishers' algorithms use hand-crafted iEEG features extracted in time or frequency domain, together with common classification models such as support vector machine (SVM), random forest (RF), k-nearest neighbor (kNN), and neural networks \cite{Brinkmann2016}.


For the above-mentioned algorithms, not only is the best combination of features and classifiers not known for every patient, but also an optimal feature set and classifier may be sub-optimal in the future because of the dynamic changes in the brain. To address these limitations, algorithms based on deep learning have been developed and tested on differentiating preictal iEEG activities from interictal ones. For instance, in \cite{Truong2018}, Truong \textit{et al.} developed a CNN architecture with three convolutional blocks and used the short-time Fourier transforms of the EEG signals as inputs to this network. This algorithm resulted in an average seizure prediction sensitivity of 75\% and a false positive rate of 0.21/h. In \cite{Eberlein2018}, the time-series iEEG signals were used as inputs to one-dimensional CNN architecture, achieving satisfactory seizure prediction performance of 0.841 AUC. In \cite{Korshunova2018}, a set of temporal and spectral iEEG features were used as inputs to a CNN framework yielded an average AUC score of 0.76. Moreover, the authors of \cite{Liu2019} designed a multi-view CNN architecture that incorporates time and frequency domain EEG features as complementary inputs. This architecture produced average AUC scores of 0.837 and 0.842 on the public and private test sets, respectively.

\subsection{Seizure prediction using human invasive EEG}

\textbf{Dataset:} Melbourne University AES/MathWorks/NIH Seizure Prediction Challenge \cite{Kuhlmann2018} --- iEEG was recorded for a long period of time (6-36 months) from 15 patients with refractory focal epilepsy. The data were collected at 400 Hz using an ambulatory seizure advisory system with 16 subdural electrodes. For the Kaggle contest, the iEEG data from three patients (Patients 3, 9 and 11) who showed lowest seizure prediction results in \cite{Cook2013prediction} were chosen. Similar to the dataset in Section \ref{sec2-1}, all iEEG data were provided in 10-minute data clips: preictal clips were extracted in six 10-minute epochs between 60 minutes and 5 minutes before a seizure, and six 10-minute interictal clips were selected in groups beginning from a randomly selected time with a minimum gap of 3 hours before and 4 hours after any seizure. Patient 1 (female, age 22), Patient 2 (female, age 51), and Patient 3 (female, age 53) provided the iEEG data spanning 559, 393, and 374 days, respectively.

\textbf{Related Work} --- Over the past decade, researchers have made several attempts to develop accurate seizure prediction methods, especially for people with drug-resistant epilepsy. The achieved prediction performance, however, was limited by the lack of long-term preictal and interictal iEEG recordings. In 2013, Cook \textit{et al.} designed a seizure advisory system (SAS) that can predict the probability of seizure occurrence in minutes or hours in advance \cite{Cook2013prediction}. The seizure prediction algorithm used by the SAS yielded prediction performances with a large subject variability. The highest seizure prediction sensitivity, for example, was 100\% for two of the patients, while it was as low as 17-45\% for other patients. 
Four years later, Cook, together with other collaborators, proposed a circadian method that maintains adequate seizure prediction performance across all patients \cite{Karoly2017}. This method yielded average sensitivity scores of 62.1\% for all the patients and 52.7\% for the three patients under study.

In late 2016, a subset of iEEG data recorded from the three patients with the lowest seizure prediction results was made publicly available via the Melbourne University Seizure Prediction Competition on Kaggle.com. Among the 478 algorithms submitted, the top-performing algorithm scored average AUC scores of 0.853 and 0.807 on the public and private test sets, respectively. This algorithm used a combination of 11 machine learning models including extreme gradient boosting, kNN, generalized linear model, and SVM along with various linear and non-linear EEG features extracted from time and frequency domains \cite{Kuhlmann2018}. The other top four performing algorithms achieved an average AUC score in the range of 0.783-0.853 and 0.746-0.799 when applied to the public and private test sets, respectively. 

The above-mentioned algorithms employed several domain-based features together with a wide range of traditional classification models (e.g., SVM, RF, gradient boosting, etc.). The engineered and hand-crafted features, however, are very prone to inter- and intra-subject variations, and hence less practical in clinical and real-world settings. This motivated researchers to use deep neural networks for building automated seizure prediction solutions. In \cite{Kiral-Kornek2018}, for example, a deep CNN framework was proposed to forecast impending seizure attacks using EEG spectrograms. This algorithm achieved superior seizure prediction results compared to those of \cite{Cook2013prediction} and \cite{Karoly2017}. Also, Reuben \textit{et al.} used multilayer perceptron neural networks with the preictal probabilities of the top Kaggle teams and yielded an average AUC score of 0.815 on the public test set \cite{Reuben2019}.



\section{Methodology}
Convolutional neural networks have been found successful in predicting epileptic seizures \cite{Truong2018, Zhang2019, Eberlein2018, Korshunova2018, Liu2019}. In this work, we propose a more efficient CNN architecture based on a new module named ``semi-dilated convolution''. The way we pre-process the data and transform it into a form appropriate for our network architecture is described below.  

\subsection{Data Pre-processing}
\textbf{Data Segmentation} --- We first divide the iEEG clips into shorter segments with the purpose of increasing the number of labeled data samples and thus improving the performance and ability of the proposed model to generalize. In this study, both canine and human iEEG clips are divided into smaller non-overlapping segments, each of which is 30-second long.   

  

\begin{figure*}[!t]
	\centering
	\includegraphics[width=0.9\linewidth]{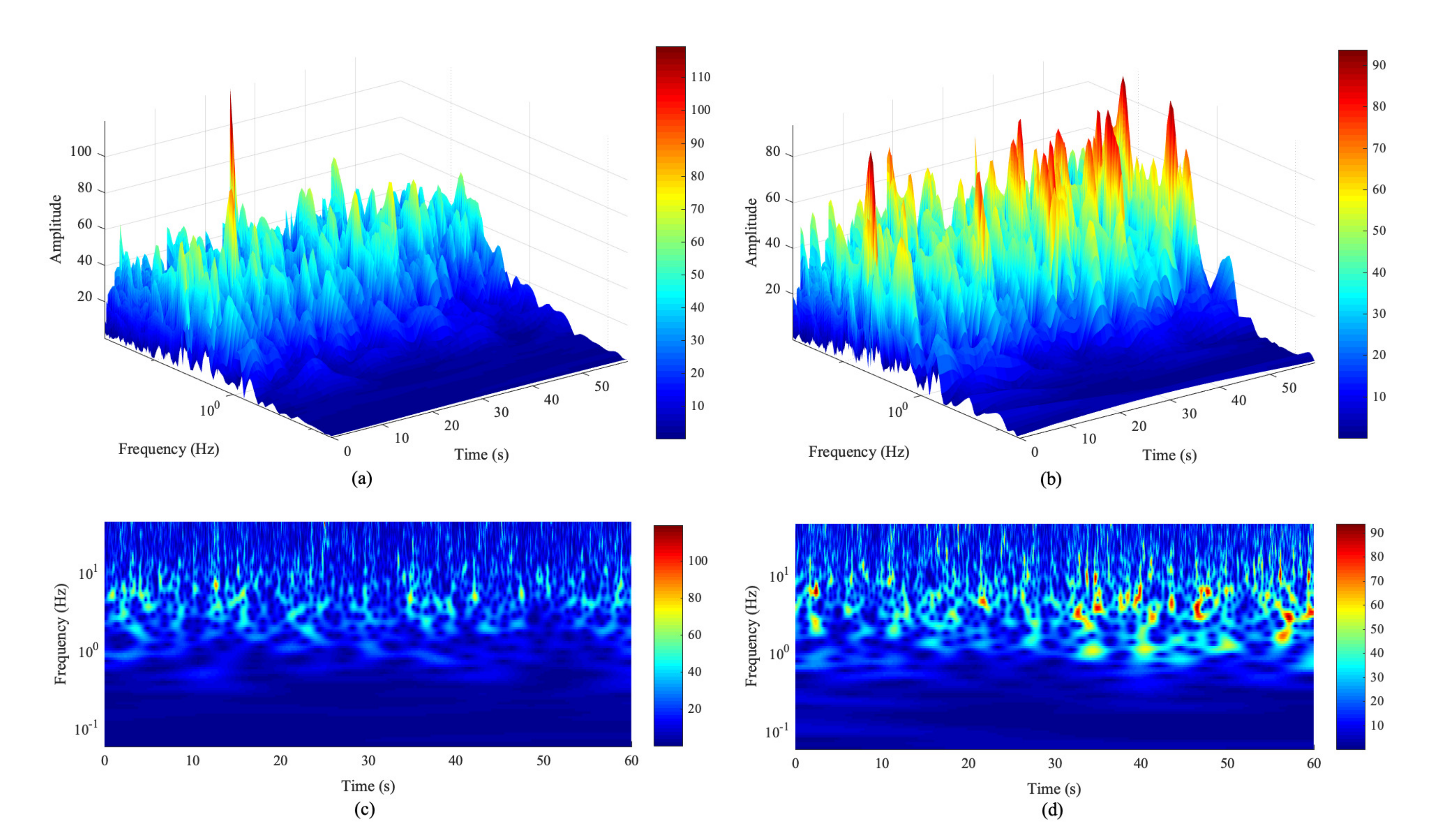}
	\caption{EEG scalogram: of time-series interictal and preictal iEEGs: (a)-(b) Scalograms of interictal and preictal iEEGs in 3D space; and (c)-(d) Scalograms of interictal and preictal iEEGs in 2D time-frequency domain.}
	\label{Fig:Scalogram}
\end{figure*}

\textbf{Data Transformation} --- Since CNNs have been proven to be a promising tool in image recognition and classification tasks,  
this inspired researchers to transform the time-series iEEG signals into image-like formats. For instance, in \cite{Truong2018}, Truong \textit{et al.} used the short-time Fourier transform to convert the 1-dimensional time-series iEEG signal into a 2-dimensional matrix with the time on the horizontal axis and frequency on the vertical axis. This transformation, however, has limited time-frequency resolution capability, due to the uncertainty principle \cite{zhang2003comparison}. On the other hand, the wavelet transform allows the use of longer time intervals where more precise low-frequency information is needed and shorter regions where high-frequency information is needed \cite{grossmann1990reading}. We, thereby, decided to apply continuous wavelet transform to the iEEG segments to produce a more faithful representation of iEEG data in the time-frequency domain. The absolute value of the CWT of a signal, as a function of time and frequency, is named a ``scalogram'' \cite{akay1998time}. Figure~\ref{Fig:Scalogram} shows the wavelet scalograms of interictal and preictal iEEG segments. It can be noticed that preictal iEEG data has more signal power than the interictal one in the frequency range of 0-50 Hz. The resulting dimension of the data is $H\times W\times N_{ch}$; where $H$ is the height of the scalogram image (i.e., number of frequency components), $W$ is the width of the scalogram image (number of time-steps in an iEEG segment), $N_{ch}$ is the number of iEEG channels. For example, $H$, $W$, and $N_{ch}$ for the Melbourne University iEEG dataset take the values of 100, 6000, and 16, respectively.   

The key challenge was how to handle the rectangular-shaped scalogram images of 100$\times$6000 dimensions using square kernels. This motivated us to develop nonsquare kernels based on a novel convolutional module named ``semi-dilated convolution''. Section~\ref{subsec:SDC} explains how semi-dilated convolutions can create a flexible-size receptive field and thus accommodates a wide range of input image shapes including EEG scalograms. Section~\ref{subsec:SDCN} describes our proposed semi-dilated convolutional network architecture (SDCN) and how it could effectively distinguish preictal iEEGs from interictal ones.  


\subsection{Semi-Dilated Convolution}
\label{subsec:SDC}


Large receptive fields in convolutional neural networks are often desired for processing large/high-resolution images (\textit{e.g.}, medical images). They, however, entail the use of large-size kernels. The drawback of this approach is it necessitates a substantial increase in the number of network parameters and run-time, making it difficult to avoid overfitting. To address these limitations, Yu and Koltun proposed to use a new convolution module named ``dilated convolution'' or ``atrous convolution'' \cite{yu2015multi}. The idea of convolution with a dilated filter was first presented in \cite{shensa1992discrete} for signal decomposition using a discrete wavelet transform. In \cite{yu2015multi}, Yu and Koltun reused the same concept to build a deep CNN architecture for multi-scale context aggregation. Their architecture, named ``context module'', had seven layers that apply 3$\times$3 convolutions with different dilation rates.  
The dilated convolution modules support exponential growth of receptive fields in both image dimensions (height and width), which may be sub-optimal for nonsquare-shape images like scalograms.  

\begin{figure*}[ht]
	\begin{center}
		\includegraphics[width=0.9\linewidth]{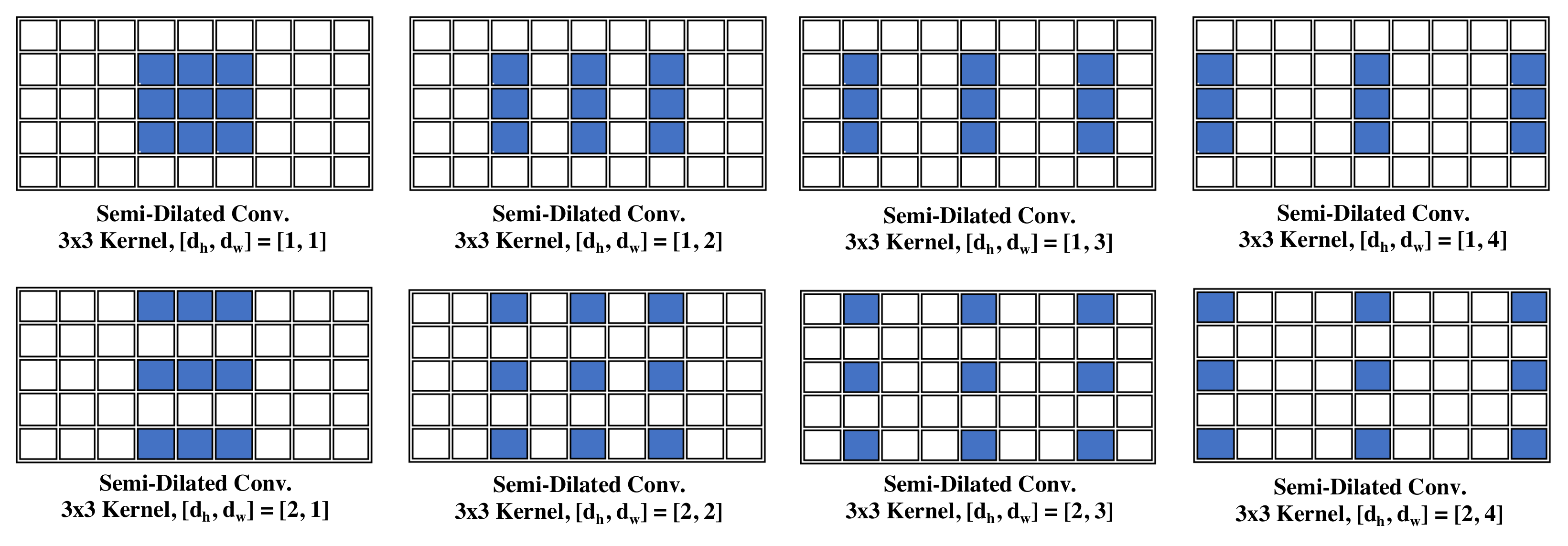}
	\end{center}
	\caption{Semi-dilated convolution with 3x3 kernel and different dilation vectors [$d_h$, $d_w$]. $d_h$ and $d_w$ correspond to the vertical dilation (along the image \textbf{h}eight) and horizontal dilation (along the image \textbf{w}idth). Top row: 3x3 semi-dilated convolutions with a vertical dilation rate of 1 and different horizontal dilation rates of 1, 2, 3, and 4, respectively. Bottom row: 3x3 semi-dilated convolutions with a vertical dilation rate of 2 and different horizontal dilation rates of 1, 2, 3, and 4, respectively.}
	\label{fig:SDC}
\end{figure*}

In this work, we propose a new convolution module named ``semi-dilated convolution'' that creates a flexible-size receptive field, which helps accommodate a wide range of image shapes. This module is specifically designed to leverage the geometry of images under study, especially those of non-uniform dimensions (e.g., 100$\times$6000 scalograms with high width to height ratio). Figure~\ref{fig:SDC} explains how the semi-dilated convolution module is different from the systematic dilated convolution and standard convolution. The dilation vector has two elements, $d_h$, $d_w$, which controls how much the receptive field will be expanded over the image height and image width, respectively. The top row of Figure~\ref{fig:SDC} displays how to apply 3$\times$3 convolution with a fixed vertical dilation rate ($d_h$) of 1 and different horizontal dilation rates ($d_w$) of 1, 2, 3, and 4. Similarly, the bottom row shows 3$\times$3 convolutions with a fixed $d_h$ of 2 and different $d_w$ of 1, 2, 3, and 4. It also clearly depicts how the shape of the receptive field differs when the vertical or horizontal dilation rate changes.

In semi-dilated convolution with $k$$\times$$k$ kernel, the receptive field of each element takes the shape of:
\[
\Big( k + (d_h-1)(k-1) \Big) \times \Big( k + \left(d_w-1\right) \left(k-1\right) \Big)
\]
\[
= \Big( \left(k - 1 \right) d_h + 1  \Big) \times \Big( \left(k-1\right) d_w + 1 \Big)
\]
where $d_h$ and $d_w$ are the vertical and horizontal semi-dilation rates, respectively. For example, for a semi-dilated convolution with 5$\times$5 kernel and a dilation vector [$d_h$, $d_w$] of [2, 16], the receptive field will take the rectangle shape of (4~$\cdot$~2 + 1)$\times$(4~$\cdot$~16 + 1) $=$ 9$\times$65.

\begin{figure*}[ht!]
	\begin{center}
		\includegraphics[width=0.9\linewidth]{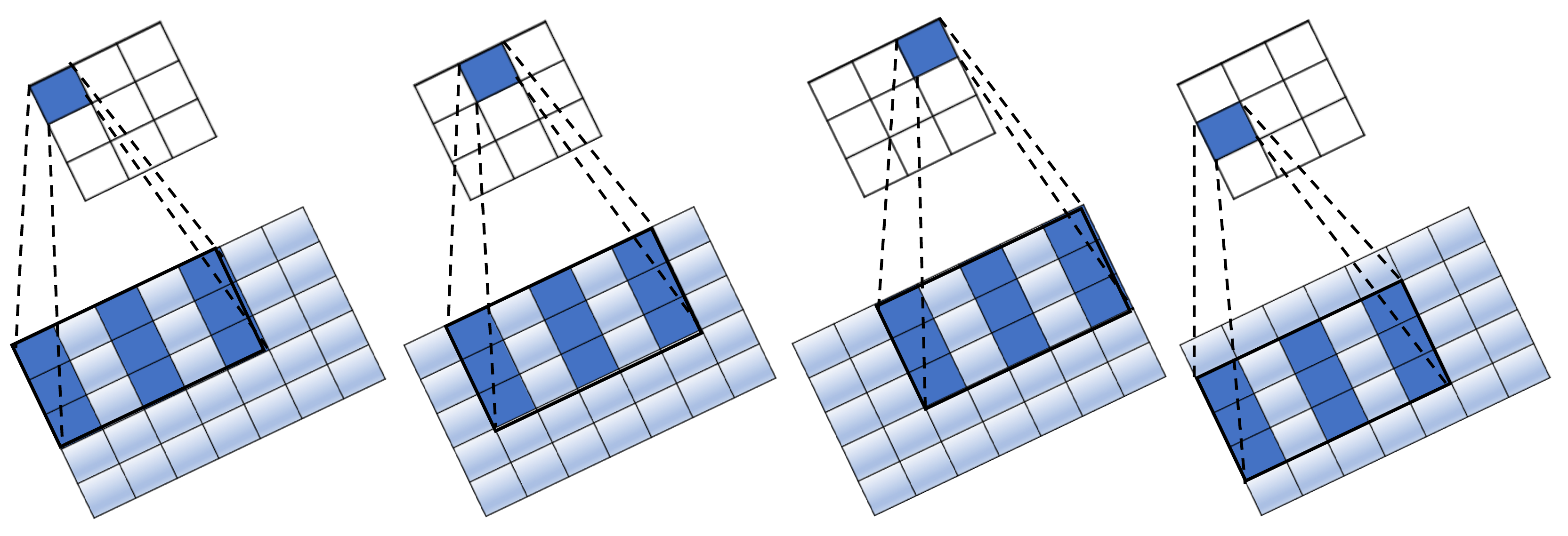}
	\end{center}
	\caption{Illustration of semi-dilated convolution: Convolving a 3x3 kernel over a 5x7 rectangular image with a dilation vector [$d_h$, $d_w$] of [1, 2].}
	\label{fig:SDN_HowItWorks}
\end{figure*}

Figure~\ref{fig:SDN_HowItWorks} also illustrates an example of semi-dilated convolution when [$d_h$, $d_w$] = [1, 2]. We can see that, unlike the dilated convolution, the receptive field of semi-dilated convolution could be more expanded over the long image dimension than over the short dimension. This helps optimize the number of parameters of the convolutional neural network without loss of resolution or coverage.


\subsection{Network Architecture}
\label{subsec:SDCN}

\begin{figure}[ht!]
	\begin{center}
		\includegraphics[width=1.00\linewidth, height=1.4\linewidth]{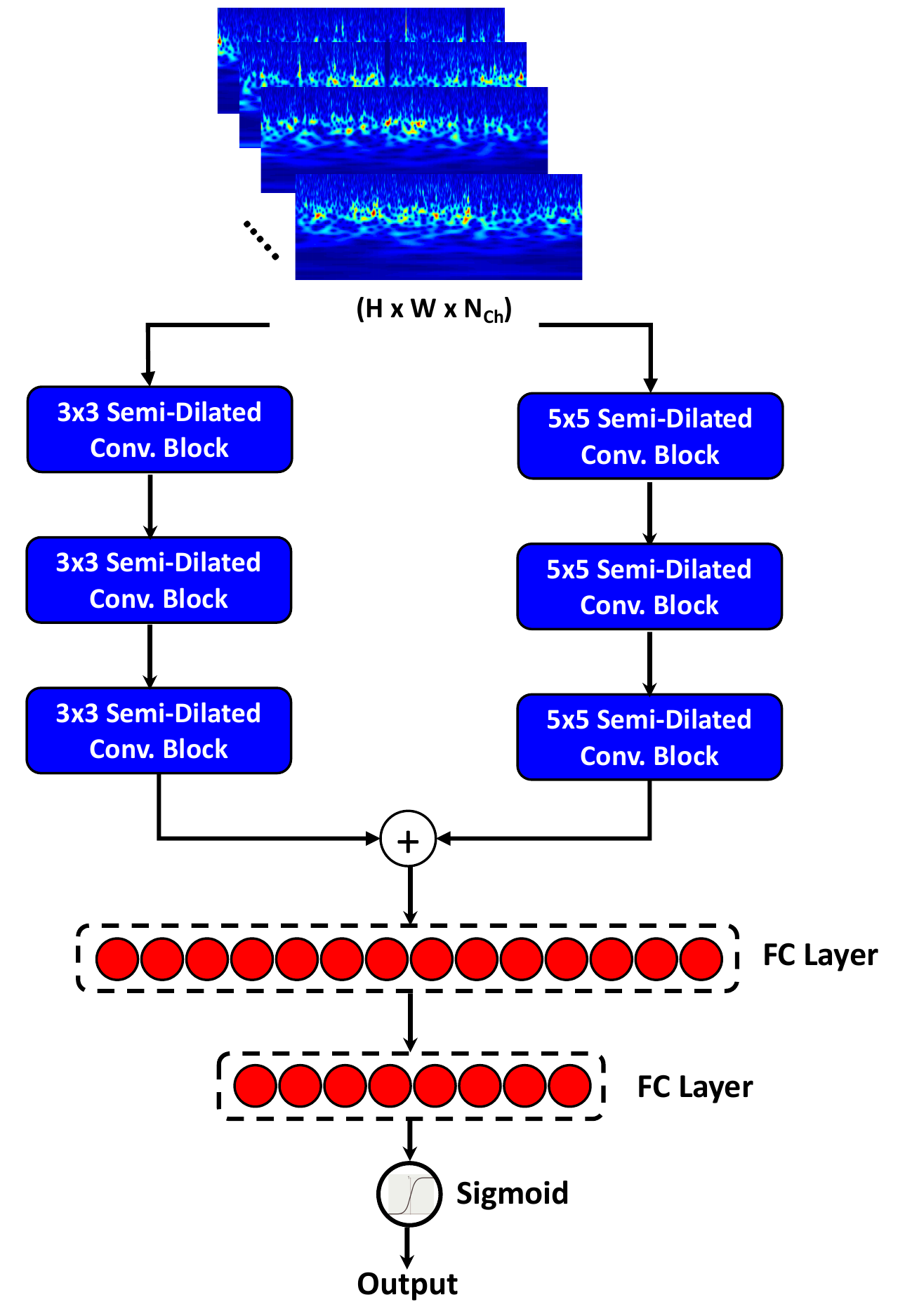}
	\end{center}
	\caption{The architecture of the proposed semi-dilated convolutional neural network: $H$ and $W$ are the height and width of the iEEG scalograms; $N_{Ch}$ is the number of iEEG channels; FC stands for Fully Connected network; and Sigmoid function is to predict the iEEG class  probabilities. 
	}
	\label{fig:SDCNetwork}
\end{figure}

In this work, we build a multi-scale CNN architecture that comprises a set of semi-dilated convolutional layers to extract and aggregate different representation maps from different scales and receptive fields. Figure~\ref{fig:SDCNetwork} depicts the schematic diagram of the proposed semi-dilated convolutional network (SDCN) architecture used for seizure prediction. The iEEG scalograms (images) are first fed into the SDCN for automated feature learning and classification. The number of channels ($N_{ch}$) is corresponding to the number of iEEG electrodes used for data acquisition, and each scalogram takes the shape of $H$$\times$$W$, where $H$ and $W$ are the scalogram height and width, respectively. Inspired by the inception model, the SDCN framework is composed of two parallel paths of different kernel sizes. The iEEG scalograms are supplied into each path separately, to extract both local features (through smaller convolutions) and high-abstracted features (through larger convolutions) simultaneously.

\begin{figure}[t]
	\begin{center}
		\includegraphics[width=1.0\columnwidth]{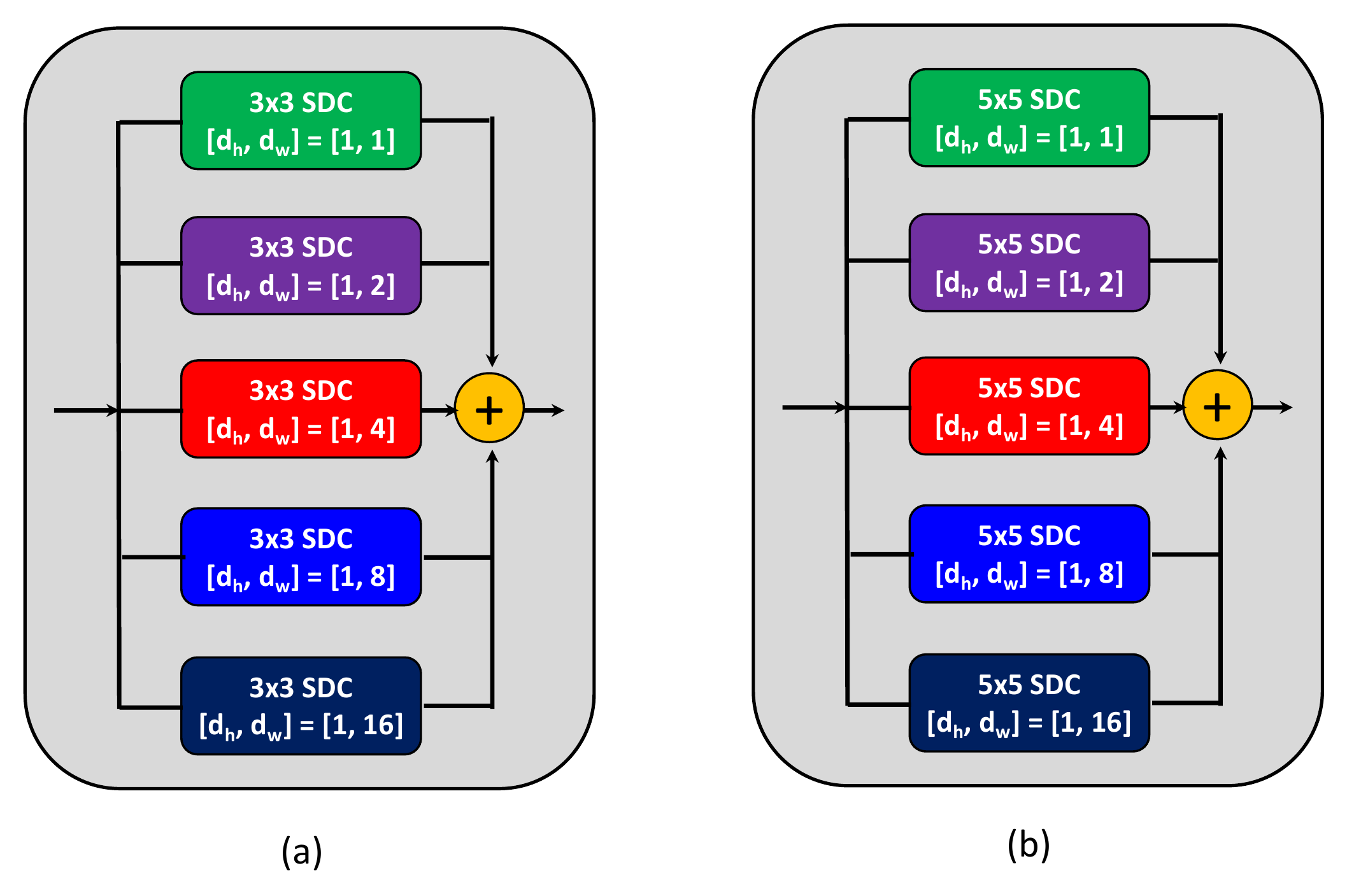}
	\end{center}
	\caption{The architecture of semi-dilated convolution blocks: (a) and (b) 3x3 and 5x5 semi-dilated convolution blocks, each contains five parallel semi-dilated convolutions with different dilation vectors.}
	\label{fig:DilationBlock}
\end{figure}

The first path consists three sequential layers, each layer has a 3$\times$3 semi-dilated convolutional (SDC) block, which is illustrated in Figure~\ref{fig:DilationBlock}. Each semi-dilated convolution block   comprises a stack of five parallel semi-dilated convolutional layers with dilation vectors [$d_h$, $d_w$]~= [1, 1], [1, 2], [1, 4], [1, 8], and [1, 16], respectively. Employing such semi-dilated convolutions with different dilation vectors is advantageous to extract and aggregate multi-view visual information without loss of resolution or coverage. A max-pooling layer of 2$\times$2 kernel is used on top of every convolution layer. The number of kernels used for the three semi-dilated convolution blocks were set to 64, 128, and 256, respectively. 

The second path has the same structure as the first path but with 5$\times$5 kernels instead. This helps extract different representation maps from both interictal and preictal iEEG scalograms, and thereby improve the performance and ability of the proposed SCDN model to generalize. The outputs of both paths are aggregated in a single feature tensor and then fed into two fully connected (FC) layers of 1024 and 512 units, respectively. Finally, a sigmoid function is used to compute the class probabilities and predictions. Since the 10-minute iEEG clips are divided into 30-second non-overlapping segments, the reported seizure prediction performance is estimated using the maximum value of the class probabilities in the 20 iEEG segments. 
Other studies use the mean value of the classification results, which has been found inaccurate for seizure prediction studies. From the network configuration perspective, our CNN model was trained by optimizing the ``binary cross-entropy'' cost function with the ``Adam'' parameter update and a learning rate of 0.001.

\section{Results and Discussion}

In this section, we examine the prediction performance of the proposed SDCN architecture and compare it to the state-of-the-art methods when tested on the datasets of (1) the American Epilepsy Society Seizure Prediction Challenge and (2) the Melbourne University Seizure Prediction Challenge. The performance metrics of sensitivity (SENS) and the AUC were used to assess the performance of our proposed SDCN framework.

\subsection{SDCN prediction performance on canine and human iEEG}

In this section, we examine the prediction performance of our SDCN model on iEEG data recorded from five dogs and two patients. We further compare its performance to the top-5 winning solutions of Kaggle American Epilepsy Society Seizure Prediction competition and also recent deep learning methods. Table~\ref{Tab_AES} summarizes the seizure prediction results achieved by the proposed and state-of-the-art methods.

\begin{table*}[!ht]
	\caption{Benchmarking of recent seizure prediction methods and the proposed SDCN architecture: American Epilepsy Society Seizure Prediction dataset}
	\begin{adjustbox}{width=\textwidth,center} 
		\label{Tab_AES}
		\begin{tabular}{| l || c | l | c | c | c |}
			\hline
			\hline
			Authors/ & Year & iEEG Features & Classifier & SENS & AUC Score  \\
			
			Team &  &  &  & (\%) &  Public/Private\\
			\hline
			Medrr \cite{Brinkmann2016} & 2016 & N/A & N/A & - & 0.903/0.840 \\
			\hline
			QMSDP \cite{Brinkmann2016} & 2016 & Spectral entropy, correlation, & LassoGLM, & - & 0.859/0.820 \\
			&  & fractal dimensions, & Bagged SVM, &  &  \\
			&  & Hurst exponent & Random Forest &  &  \\
			\hline
			Birchwood \cite{Brinkmann2016} & 2016 & Log spectral power, covariance & SVM & - & 0.839/0.801 \\
			\hline
			ESAI CEU-UCH \cite{Brinkmann2016} & 2016 & Spectral power, correlation & Neural Network, & - & 0.825/0.793 \\
			&  & derivative PCA, ICA preprocessing & kNN &  &  \\
			\hline
			Michael Hills \cite{Brinkmann2016} & 2016 & Spectral power, correlation & SVM & - & 0.862/0.793 \\
			&  & spectral entropy, fractal dimensions &  &  & \\
			\hline
			Truong \textit{et al.} \cite{Truong2018} & 2018 & Spectrogram & CNN & 75.0 & -  \\
			\hline
			Eberlein \textit{et al.} \cite{Eberlein2018} & 2018 & Multi-channel time series & CNN & - & 0.841/- \\
			\hline
			
			
			Korshunova \textit{et al.} \cite{Korshunova2018} & 2018 & Time series of spectral power & CNN & - & 0.780/0.760 \\
			\hline
			Liu \textit{et al.} \cite{Liu2019} & 2019 & Time series (PCA), spectral power & Multi-view CNN & - & 0.837/0.842 \\
			\hline
			\textbf{Proposed method} & 2020 & Continuous wavelet transform & SDCN & \textbf{88.45} & \textbf{0.928/0.856} \\
			\hline
			\hline
		\end{tabular}
	\end{adjustbox}
\end{table*}

In \cite{Brinkmann2016}, Brinkmann \textit{et al.} describes the algorithms developed by the top 10 Kaggle winning teams (we report the top 5 only in Table~\ref{Tab_AES} for the sake of brevity). All of these seizure prediction algorithms rely on a large set of hand-crafted measures such as spectral entropy, correlation, fractal dimensions, Hurst exponent, spectral power, covariance, and bag-of-wave features. These domain-based feature engineering techniques are usually unreliable because of the dynamic changes of the brain across different patients and over time for the same patient.
The top-performing five solutions' AUC scores are in the range of 0.825-0.903 on the public leaderboard (\textit{i.e.}, public test set) and 0.793-0.840 on the private leaderboard (\textit{i.e.}, private test set) \cite{Brinkmann2016}. Our proposed seizure prediction algorithm, on the other hand, use semi-dilated convolutions for automatic feature learning and classification of preictal and interictal iEEG activities. A stack of semi-dilated convolutions with different kernel sizes and different semi-dilation rates are applied to the iEEG scalograms to extract the distinguishable feature maps needed for precise prediction of seizures. This helps achieve superior AUC scores of 0.928 and 0.856 on the public and private test data sets, respectively.          

We also demonstrate a benchmark of recent deep learning-based seizure prediction methods and this work. For example, in \cite{Truong2018}, Truong \textit{et al.} used a nine-layers CNN model together with the iEEG spectrograms to identify the changes in the time-frequency domain between interictal and preictal iEEG activities. This model achieved a reasonable seizure prediction sensitivity of 75.00\%.
Our proposed SDCN structure outperforms Truong's CNN model by a significant margin; producing an average seizure prediction sensitivity of 88.45\%. 

Another elegant CNN-based seizure prediction approach was introduced in \cite{Eberlein2018}, where one-dimensional convolutions were applied to the multi-channel time-series iEEG signals without any pre-processing or data transformation. The obtained AUC score on the public test set was 0.841. In \cite{Korshunova2018} Korshunova \textit{et al.} also used a combination of time-domain and frequency-domain features achieving inferior seizure prediction AUC scores of 0.78 and 0.76 on the public and private leaderboard test sets, respectively. Further, the authors of \cite{Liu2019} designed an accurate seizure prediction algorithm that takes time-series iEEG data and their corresponding frequency spectra as inputs for a multi-view CNN framework, achieving the highest AUC score of 0.842 on the private test set. Notably, our SDCN algorithm yields superior seizure prediction AUC scores of 0.928 and 0.856 on the public and private test sets, respectively. This high performance demonstrates how our seizure predictor can effectively accommodate the variations in iEEG data across different subjects and also over time for the same subject.

\begin{table*}[!ht]
	\caption{Benchmarking of recent seizure prediction methods and the proposed SDCN architecture: Melbourne University AES/MathWorks/NIH Seizure Prediction dataset}
	\begin{adjustbox}{width=\textwidth,center} 
		\label{Tab_Melbourne}
		\begin{tabular}{| l || c | l | c | c | c |}
			\hline
			\hline
			Authors/Team & Year & iEEG Features & Classifier & SENS & AUC Score \\
			
			&  &  &  & (\%) & Public/Private \\
			\hline
			
			Cook \textit{et al.} \cite{Cook2013prediction} & 2013 & Signal energy & Decision tree, kNN & 33.67 & - \\
			\hline
			Karoly \textit{et al.} \cite{Karoly2017} & 2017 & Signal energy, circadian profile & Logistic regression & 52.67 & - \\
			\hline
			Kiral-Kornek \textit{et al.} \cite{Kiral-Kornek2018} & 2018 & Spectrogram, circadian profile & CNN & 77.36 & - \\
			\hline
			Team A  & 2018 & Spectral power, distribution statistics, & Extreme gradient  & - & 0.853/0.807 \\
			
			(1st place) \cite{Kuhlmann2018} &  & AR error, fractal dimensions, Hurst exponent, & boosting, &  &  \\
			&  & cross-frequency coherence, other features & kNN, SVM &  &  \\
			\hline
			Team B & 2018 & Correlation, distribution statistics, entropy, & Extremely & - & 0.783/0.799 \\
			(2nd place) \cite{Kuhlmann2018} &  & zero crossings, spectral energy, maximum & randomized trees &  & \\
			& & frequency, other features &  & \\
			\hline
			Team C  & 2018 & Spectral power, distribution statistics,  & SVM, random under- & - & 0.815/0.797 \\
			(3rd place) \cite{Kuhlmann2018} &  & RMS of signal, correlation, spectral edge, & sampling boosted &  & \\
			&  & first and second derivatives & tree ensemble &  &  \\
			\hline
			Team D  & 2018 & Spectral power, correlation, spectral & Gradient boosting, & - & 0.854/0.791 \\
			(4th place) \cite{Kuhlmann2018} &  & entropy, spectral edge power & SVM &  & \\
			\hline
			Team E  & 2018 & Spectral power, correlation, spectral entropy & Adaptive boosting, & - & 0.844/0.746 \\
			(5th place) \cite{Kuhlmann2018} &  & Shannon entropy, spectral edge frequency & gradient boosting, &  & \\
			& & Hjorth parameters, fractal dimensions & random forest  &  & \\
			\hline
			Reuben \textit{et al.} \cite{Reuben2019} & 2019 & Preictal probabilities from the top & Multilayer perceptron & - & 0.815/- \\
			&  & 8 teams in \cite{Kuhlmann2018} & neural network &  & \\
			\hline
			\textbf{Proposed method} & 2020 & Continuous wavelet transform & SDCN & \textbf{89.52} & \textbf{0.883/-} \\
			\hline
			\hline
		\end{tabular}
	\end{adjustbox}
	\raggedright
	\justify 
	Patient~1, 2, and 3 in the Kaggle compeition dataset are the same as Patients~3, 9, and 11 in \cite{Cook2013prediction}, \cite{Karoly2017}, and \cite{Kiral-Kornek2018}.
\end{table*}

\subsection{SDCN prediction performance on human iEEG}

We compared the results from our SDCN model with those of the recent studies (\cite{Cook2013prediction}, \cite{Karoly2017}, \cite{Kiral-Kornek2018}, \cite{Kuhlmann2018}A-E, and \cite{Reuben2019}) that were tested on the same human iEEG dataset. In \cite{Cook2013prediction}, satisfactory seizure prediction results were achieved for most of the patients (average sensitivity of 61.2\%) while the three patients under study showed prediction sensitivities of 33.7\% on average. The main reason behind this performance decay was the temporal drift (variation) observed in the adopted time-dependent iEEG features (this study did not report what iEEG features were extracted and what classifier was used to evaluate their effectiveness). In \cite{Karoly2017}, the same authors proposed a circadian seizure forecasting method to improve the prediction performance across all patients. A significant improvement in the seizure prediction was made by exploiting this circadian information. Using a logistic regression classifier, an average prediction sensitivity score of 62.1\% was achieved for all patients and 52.7\% was achieved for the three patients under study.

In \cite{Kiral-Kornek2018}, convolutional neural networks were used to boost the seizure prediction performance while relaxing the need of data pre-processing. The iEEG data clips were segmented and then transformed into time-frequency representations named spectrograms. These spectrograms were used as inputs for a CNN framework that distinguishes between preictal and interictal brain states with an average prediction sensitivity of 77.36\%. It is worth noting that the seizure prediction algorithms presented in \cite{Cook2013prediction, Karoly2017, Kiral-Kornek2018} could not achieve adequate prediction performance for Patients 3, 9, and 11 (same patients as in our study). Our semi-dilated CNN architecture, on the other hand, can accurately recognize pre-seizure patterns hidden in the preictal iEEG data, achieving an average prediction sensitivity of 89.52\%. Table~\ref{Tab_Melbourne} reports the prediction results achieved by the proposed and baseline seizure prediction methods.

Also, we compare our results to those of the winning solutions in the 2016 Kaggle seizure prediction challenge \cite{Kuhlmann2018}. The winning team deployed 11 different classification models and more than 3000 hand-crafted iEEG features and achieved average AUC scores of 0.853 and 0.807 on the public and private test sets, respectively  (see Table~\ref{Tab_Melbourne}). It is, however, impractical to compute 3000+ hand-crafted features in real-time applications. Our seizure prediction algorithm remarkably yields a superior AUC score of 0.883 on the public test set without involving any computationally-intensive operations, making it more suitable for use in ambulatory and clinical applications.

\section{Conclusion}

This work proposes a novel seizure prediction algorithm by learning a semi-dilated convolutional network (SDCN). The proposed semi-dilated convolution module is advantageous to handle a wide range of image shapes as it allows performing convolutions with different vertical and horizontal dilation rates. This helps create a flexible-size receptive field that could be largely expanded over the long dimension and moderately expanded over the short dimension of the input image. Our network is applied to the scalogram images of the invasive EEG signals and tested on two large-scale and publicly available EEG datasets. Experimental results show that the proposed method significantly advances the state-of-the-art in both datasets, achieving an average seizure prediction sensitivity of 88.45-89.52\% and AUC of 0.88-0.92.
   


\section{Limitations}
The prediction performance of our SDCN architecture decays when the EEG data includes outliers or gets contaminated by any of the physiological or non-physiological artifacts. We will release our codes and trained models to support progress in this field. 





{\small
\bibliographystyle{IEEEtran}
\bibliography{refs}
}

\end{document}